\documentclass[11pt,a4paper]{article}
\usepackage{jcappub}
\usepackage{bm}
\usepackage{multirow}
\usepackage{booktabs}
\usepackage{xcolor}

\def\LJD{$\Lambda\mbox{JD16}^*$}
\def\LJDF{$\Lambda\mbox{JDF}^*$}

\begin{document}

\title{Influence of the correlation prior on reconstruction of the dark energy equation of state}

\author[a,b]{Youhua Xu}
\author[b,c]{Hu Zhan}
\author[a,d]{and Yeuk-Kwan Edna Cheung}

\affiliation[a]{School of Physics, Nanjing University, 
22 Hankou Road, Nanjing, Jiangsu 210093, China}
\affiliation[b]{CAS Key Laboratory of Space Astronomy and Technology, 
National Astronomical Observatories, Beijing 100101, China}
\affiliation[c]{Kavli Institute for Astronomy and Astrophysics, 
Peking University, Beijing 100871, China}
\affiliation[d]{Institute of Nuclear and Particle Physics,
Demokritos National Research Centre,  Athens,  Greece}

\emailAdd{yhxu@nao.cas.cn}
\emailAdd{zhanhu@nao.cas.cn}
\emailAdd{cheung@nju.edu.cn}

\arxivnumber{}

\abstract{
  Non-parametric reconstruction of the dark energy equation of
  state (EoS) aims to determine the EoS as a function of redshift
  without invoking any particular dark energy model, so that the
  resulting EoS can be free of model-induced biases or artifacts.
  Without proper regularization, however, such reconstruction is
  often overwhelmed by the noise of poorly constrained modes.
  An intuitive regularization scheme is to assume a priori the dark
  energy EoS to evolve at most slowly with time, which may be
  enforced by a correlation between the EoS at different
  epochs. Indeed, studies that impose the correlation prior
  are able to significantly reduce the uncertainties of the
  reconstructed EoS and even show hints for dynamical dark energy.
 
  In this work, we examine the correlation prior using
  mock datasets of type Ia supernovae (SNe Ia), baryonic acoustic
  oscillations (BAOs), age-derived Hubble parameter, Hubble constant,
  and cosmic microwave background.
  We find that even though the prior is designed to disfavor
  evolving equations of state, it can still accommodate spurious
  oscillating features at high significance.
  Within the 1000 mock datasets of existing observations that are
  generated for the concordance cosmological model,
  i.e., the input dark energy EoS $w=-1$, there are 688 (69)
  cases recovering an EoS that departs from $-1$ by more than
  $1\sigma$ ($2\sigma$) in one or more redshift bins.
  The reconstructed EoS turns up and down markedly in many cases.
  Moreover, inverting the signs of the randomly assigned
  errors of the mock data more or less reverses the 
  behavior of the EoS.
  Spurious results occur even more frequently when idealized SN Ia
  and BAO data from future surveys are included.
  Our tests suggest that further studies are needed to ensure
  accurate reconstruction of the EoS with the correlation prior.
}

\maketitle

\section{Introduction}
The discovery of accelerated cosmic expansion~\citep{1998AJ....116.1009R,Perlmutter1999ApJ}
has inspired widespread efforts to determine its underlying physics
(see, e.g., \citep{2008ARA&A..46..385F,2009ARNPS..59..397C,
2011CoTPh..56..525L,2012Ap&SS.342..155B,2013PhR...530...87W} for reviews).
One of the solutions proposed is to add a smooth component to
the universe -- dark energy in a broad sense -- that drives the
acceleration with its negative pressure.
The equation of state (EoS), i.e., the ratio of pressure to density
$w \equiv p/\rho$, is often used to characterize dark energy models.
Since the EoS is inferred indirectly from geometry and growth
measurements, one can define an effective dark energy
EoS for theories that do not invoke dark energy at all.
As such, the dark energy EoS can serve as a phenomenological
discriminator for different theories of the accelerated cosmic expansion.
It has indeed become a major task for next-stage dark energy surveys
such as the Large Synoptic Survey
Telescope\footnote{\url{https://www.lsst.org/}.} (LSST),
\emph{Euclid}\footnote{\url{http://sci.esa.int/euclid/}.}, 
DESI\footnote{\url{http://desi.lbl.gov/}.}, 
and \emph{WFIRST}\footnote{\url{http://wfirst.gsfc.nasa.gov/}.} to test
various theories with accurate EoS measurements over a wide range of
redshift.

Non-parametric reconstruction plays an important role in measuring the
dark energy EoS~\citep{2003PhRvL..90c1301H,2006IJMPD..15.2105S}. 
It tries to avoid potential biases or artifacts arising from specific
model assumptions by adopting as little knowledge about the EoS as possible,
though details of the implementation can still lead to different conclusions.
For example, there are reports hinting at an evolving EoS and,
hence, dynamical nature of dark energy~\citep{2012PhRvL.109q1301Z,2017NatAs...1..627Z,
2017APh....86....1Z,2018ApJ...869L...8W,2018ApJ...857....9D,2019arXiv190209794Z}, 
whereas others find consistency with the cosmological constant, i.e., $w=-1$ at
all time~\citep{2009PhRvD..80l1302S,PhysRevLett.105.241302,2013PhRvD..88d3515S,
2014PhRvD..89b3004W,2016A&A...594A..14P,2017MNRAS.466..369H,
2019arXiv190411068R}. The discrepancy calls for a closer examination of the 
reconstruction method.

In non-parametric reconstruction, the EoS is replaced by an approximation
function that is constructed from its values $w_i\equiv w(a_i)$ at a series
of scale factors $a_i$ (or redshifts $z_i$). The approximation function is
often chosen to be piece-wise constant or an interpolation of
$\{w_i\}$~\citep{2003PhRvL..90c1301H,2005PhRvD..71b3506H,2009JCAP...12..025C,
2014PhRvD..89b3004W,2012JCAP...09..020V}. The EoS values $\{w_i\}$ are then
estimated from data and can vary freely to accommodate dark energy evolution
to the extent the approximation function allows. In this way, one can be
confident that the resulting EoS is completely determined by the data rather
than some subtle effects of the particular dark energy model assumed in the
first place. However, the drastically increased degrees of freedom also cause
large uncertainties on the recovered EoS, making it less useful for
testing dark energy models. 

As pointed out in ref.~\citep{2003PhRvL..90c1301H}, the uncertainties of
the reconstructed EoS are dominated by poorly constrained eigenmodes,
which typically oscillate more rapidly than well determined eigenmodes do.
Since there is no strong case for a fast-oscillating EoS, it is
reasonable to treat these poorly constrained modes as unphysical. 
One can thus reduce the EoS uncertainties by filtering out such eigenmodes
in the result~\citep{2003PhRvL..90c1301H,2005PhRvD..71b3506H},
though the final EoS and its errors would depend on the number of eigenmodes 
being kept~\citep{2017APh....86....1Z,2018ApJ...857....9D}.
Another way to reduce the EoS uncertainties is to regularize
the reconstruction process with a correlation prior to suppress
fast-varying modes in the posterior distribution of
$\{w_i\}$~\citep{2012JCAP...02..048C}.
The correlation prior introduces an additional $\chi^2$ term into
the $\{w_i\}$ likelihood, which is derived from an assumed correlation
function of the dark energy EoS at different epochs. 
Recent applications of this method have shown preference for
dynamical dark energy over the cosmological constant at more than
$3\sigma$ significance level~\citep{2017NatAs...1..627Z}. 

Given the crucial role of the dark energy EoS in deciphering the
accelerated cosmic expansion, one must ensure that the reconstruction
method recovers the true EoS. While a thorough study of non-parametric
reconstruction is needed, such an undertaking is well beyond the
scope of this paper. Here we only test whether the correlation prior
allows signatures of dynamical dark energy to be ``reconstructed'' 
from a universe dominated by the cosmological constant and cold dark
matter (the $\Lambda$CDM model).
For convenience, we refer to the cosmological model that allows
$w$ to vary in time as the $\tilde{w}$CDM model.

The rest of the paper is organized as follows.
Section~\ref{sec:method} describes the method of non-parametric 
reconstruction of the dark energy EoS with the correlation prior
and reproduces the result in ref.~\citep{2017NatAs...1..627Z}. 
In section~\ref{sec:tests}, we test the correlation prior using
mock datasets of current observations and future surveys 
that are generated for the $\Lambda$CDM universe.
Section~\ref{sec:summary} summarizes the results and concludes
the paper.

\section{Reconstructing the dark energy EoS}
\label{sec:method}

\subsection{Non-parametric reconstruction with the correlation prior}
\label{subsec:method}

Non-parametric reconstruction of the dark energy EoS is essentially
estimating a discrete set of EoS values $\{w_i\}$ that are used to
interpolate the EoS as a function of time (or its equivalent).
The value of $w_i$ at each interpolation node is allowed
to vary freely to accommodate arbitrary evolution of dark energy.
Piece-wise constant interpolation and linear interpolation are 
common choices for approximating the EoS~\citep{2003PhRvL..90c1301H,
2005PhRvD..71b3506H,2009JCAP...12..025C,2014PhRvD..89b3004W,2012JCAP...09..020V}.
We find no significant difference between the results of the
two schemes as long as there are enough interpolation nodes to
cover the range of the scale factor of all the data.

Following refs.~\citep{2012JCAP...02..048C,2012PhRvL.109q1301Z,2017NatAs...1..627Z},
we adopt the piece-wise constant approximation.
The scale factor interval $[a_\mathrm{min},1]$ is evenly divided into $N$
bins of width $\Delta a=(1-a_{\rm min})/N$ with each bin assigned a $w_i$. 
The choice of $a_{\rm min}$ is determined by the scale factor of the
highest-redshift data point excluding the cosmic microwave background (CMB).
The distance to the last scattering surface, as derived from the CMB,
is the only data point that contains dark energy information 
at $a<a_{\rm min}$. 
Therefore, we assign a single EoS parameter to the interval
$[a_{\rm CMB},a_{\rm min}]$.

With the observables expressed in terms of $\{w_i\}$ and other
parameters, one can use Markov Chain Monte Carlo (MCMC) sampling to
map the posterior distribution
\begin{equation}
  \mathcal{P}(\bm{\theta}|\bm{d})\propto P(\bm{\theta})\,
  \mathcal{L}(\bm{d}|\bm{\theta}),
\end{equation}
where $\bm{\theta}$ is the full set of parameters including $\{w_i\}$,
$\bm{d}$ is the data vector, $P(\bm{\theta})$ is the prior, and
$\mathcal{L}(\bm{d}|\bm{\theta})$ is the likelihood of $\bm{d}$ given 
$\bm{\theta}$. The likelihood function is approximated as a multivariate
Gaussian $\mathcal{L}(\bm{d}|\bm{\theta}) \propto
\exp\left(-\frac{1}{2}\chi_{\rm data}^2\right)$ with
\begin{equation} \label{eq:chi_data}
  \chi_{\rm data}^2 = \left(\bm{d}-\bm{d}^{\rm th}\right)^{\rm T}
  \bm{C}_d^{-1} \left(\bm{d}-\bm{d}^{\rm th}\right),
\end{equation}
where $\bm{d}^{\rm th}$ is the data vector calculated theoretically from
$\bm{\theta}$, and $\bm{C}_d$ is the covariance matrix of the data.
Note that $\bm{C}_d$ may depend on $\bm{\theta}$.

To reduce the uncertainties and enforce the smoothness
of the reconstructed EoS,
ref.~\citep{2012JCAP...02..048C} proposes
to apply a prior 
$P(\bm{w}) \propto \exp(-\frac{1}{2}\chi_{\rm prior}^2)$ with
\begin{equation} \label{eqn:chi_prior}
  \chi_{\rm prior}^2 = \left[(\bm{I}-\bm{S})\bm{w}\right]^{\rm T}
  \bm{C}_w^{-1} \left[(\bm{I}-\bm{S})\bm{w}\right],
\end{equation}
where $\bm{w} \equiv \{w_i\}$, $\bm{I}$ is the identity matrix,
$\bm{S}$ is the smoothing matrix, and $\bm{C}_w$ is
the covariance matrix of $\bm{w}$. The total $\chi^2$ for the MCMC sampling is
then $\chi_{\rm tot}^2 = \chi_{\rm data}^2 + \chi_{\rm prior}^2$. 
The covariance matrix $\bm{C}_w$ is derived from an assumed
correlation function of the dark energy EoS:
\begin{eqnarray}
  C_{w,ij} &=& \frac{1}{(\Delta a)^2}\int^{a_i+\Delta a}_{a_i}da
  \int^{a_j+\Delta a}_{a_j}da' \xi_{w}(|a-a'|), \\
  \xi_{w}(|a-a'|) &\simeq& \frac{\sigma^2_{\bar w}}{\pi a_c}\,
  \frac{1-a_{\rm min}}{1+ \left(|a-a'|/a_c\right)^2},
\end{eqnarray}
where $\sigma^2_{\bar w}$ is the variance of the mean value of $w(a)$
within the interval $[a_{\rm min},1]$, and $a_c$ is the correlation length.
The smoothing matrix $\bm{S}$ performs a local average of
$\bm{w}$, i.e.,
\begin{equation}
  S_{ij}=\left\{ \begin{array}{lr}
    1/N_i & \quad |a_i - a_j| \le a_c \\
    0     & \mbox{else} \end{array} \right . ,
\end{equation}
with $N_i$ being the number of the EoS bins in the range $[a_i-a_c,a_i+a_c]$.
We adopt the same ``weak prior'' ($\sigma_{\bar w}=0.04$ and $a_c=0.06$)
used in refs.~\cite{2012PhRvL.109q1301Z,2017NatAs...1..627Z} throughout this paper.

\begin{figure}[t!]
\centering
\includegraphics[width=\textwidth]{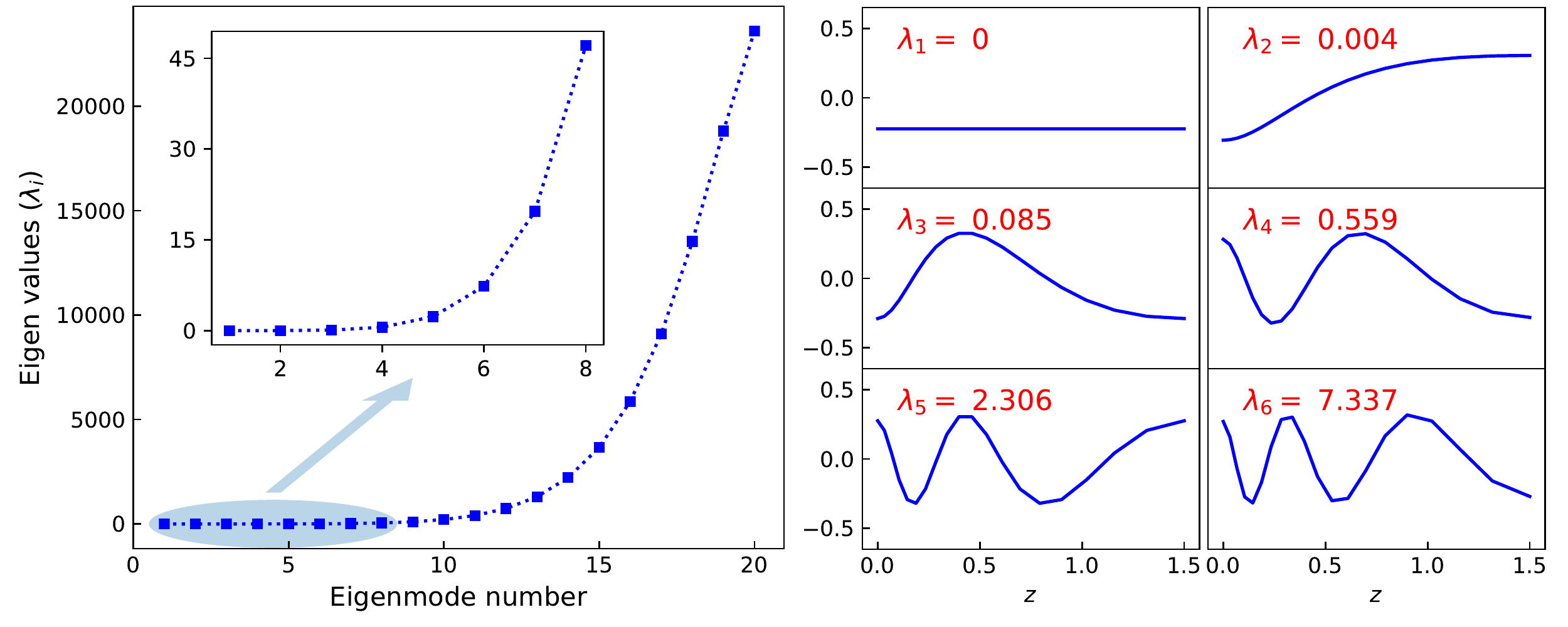}
\caption{
  \emph{Left panel}: 
  Eigenspectrum of the correlation prior with
  $\sigma_{\bar{w}}=0.04$, $a_c=0.06$, $N=20$ and $a_{\rm min}=0.4$.
  The eigenmodes are ordered by the number of zero-crossing nodes.
  The inset is a zoom-in of the first 8 modes.
  \emph{Right panel}:
  The first six eigenmodes in redshift coordinate.
}
\label{fig:eigen_spectrum}
\end{figure}

The $\chi^2$ contribution of the correlation prior, i.e.,
eq.~\eqref{eqn:chi_prior}, can be rewritten more concisely as
$\chi^2_{\rm prior}=\bm{w}^{T} \bm{F}_w \bm{w}$ with
$\bm{F}_w = (\bm{I}-\bm{S})^{\rm T} \bm{C}_w^{-1} (\bm{I}-\bm{S})$.
Projecting $\bm{w}$ onto the complete set of eigenmodes of
$\bm{F}_w$, one obtains 
\begin{equation} \label{eqn:chi2_p_em}
  \chi^2_{\rm prior} =
  \sum_{i=1}^{N} \lambda_i \left(\bm{w}^{\rm T} \bm{e}_i\right)^2,
\end{equation}
where $\bm{e}_i$ is the $i$-th eigenmode of $\bm{F}_w$,
and $\lambda_i$ is its corresponding eigenvalue.
Figure~\ref{fig:eigen_spectrum} gives an example of the eigenspectrum
of the correlation prior. The eigenmodes are ordered by the number
of zero-crossing nodes: the higher the mode number, the more the
eigenmode oscillates. The eigenvalue increases quickly with the mode
number, penalizing fast-varying modes through
eq.~\eqref{eqn:chi2_p_em}. As expected, the $\chi^2$ contribution of
the correlation prior vanishes for the constant EoS.

\subsection{Results of existing observations}
\label{subsec:checks}

We implement a cosmological parameter estimation program
ClassMC\footnote{\url{https://github.com/xyh-cosmo/ClassMC}.}
to reconstruct the dark energy EoS, which incorporates
the Boltzmann code
CLASS~\citep[][v2.5.0]{2011arXiv1104.2932L,2011JCAP...07..034B}
(modified to allow the dark energy EoS to evolve arbitrarily)
and the affine-invariant ensemble sampler~\citep{2010CAMCS...5...65G}
IMCMC\footnote{\url{https://github.com/xyh-cosmo/imcmc}.}~\citep{2018JCAP...03..045X}.
The sampled chains are analyzed using 
GetDist\footnote{\url{https://github.com/cmbant/getdist}.}.
For simplicity, we assume a flat geometry and ignore dark energy
clustering~\citep{2005PhRvD..72l3515Z,2011PhLB..702....5L}, which are
not expected to alter the conclusion of this study.

We use the following observations to reconstruct the dark energy EoS: 
\emph{Planck} CMB temperature and polarization power
spectra\footnote{Likelihood products: {plik\_lite\_v18\_TTTEEE.clik}
  and {lowl\_SMW\_70\_dx11d\_2014\_10\_03\_v5c\_Ap.clik}.}~\citep[]
[\emph{Planck}-2015]{2016A&A...594A...1P,2016A&A...594A..11P},
SDSS-II/SNLS3 type Ia supernova (SN Ia) joint light-curve analysis 
data~\citep[][JLA]{2013A&A...552A.124B,2014A&A...568A..22B},
BOSS DR12 baryon acoustic oscillation (BAO) measurements of the angular
diameter distance $D_A(z)$ and the Hubble parameter $H(z)$ in
combination with the comoving sound horizon at the drag epoch
$r_d$~\citep{2017MNRAS.466..762Z,2017MNRAS.469.3762W},
the Hubble parameter $H(z)$ derived from galaxy
ages~\citep[][OHD]{2016JCAP...05..014M}, the Hubble constant
$H_0=73.24\pm 1.74~\rm{km~s^{-1} Mpc^{-1}}$~\citep{2016ApJ...826...56R},
BAO $D_A(z)$ and $H(z)$ measurements from the Ly$\alpha$
forest~\citep{2015A&A...574A..59D}, as well as BAO $D_V(z)$
measurements from the 6dF galaxy survey~\citep{2011MNRAS.416.3017B}
and the SDSS DR7 Main Galaxy Sample~\citep{2015MNRAS.449..835R}.
This dataset includes all the data of the ALL16 combination in
ref.~\citep{2017NatAs...1..627Z} except the WiggleZ galaxy power
spectra~\citep{2012PhRvD..86j3518P} and the CFHTLenS weak-lensing
shear power spectra~\citep{2013MNRAS.432.2433H}, which do not have 
a significant impact on the reconstructed EoS.
The data of CMB, SN Ia, BAO, OHD, $H_0$, and Ly$\alpha$ forest
are assumed to be mutually independent, so that the total covariance
matrix $\bm{C}_d$ has a block-diagonal form with one block for each
type of observation.
Moreover, to keep the computational cost manageable, we replace the
CMB power spectra with the CMB distance prior~\citep{2015JCAP...12..022H}
for the ensemble study in section~\ref{sec:tests}, and, hence,
a consistency check between the results of the two types of CMB data is
performed in this section. We refer to the joint dataset with the CMB
power spectra as JD16 and that with the CMB distance prior as JD16$^*$.

The set of parameters to be estimated is
\begin{eqnarray} \label{full_param_space}
  \bm{\theta} \equiv \left\{H_0,~\Omega_b,~\Omega_c,~A_s,~n_s,~\tau,~w_0,
  \dots,w_{29},~\mathcal{N} \right\},
\end{eqnarray}
where $\Omega_b$ is the baryon density parameter, $\Omega_c$ is the
cold dark matter density parameter, $A_s$ and $n_s$ are, respectively,
the amplitude and the spectral index of the primordial density power
spectrum, $\tau$ is the optical depth due to reionization, and
$\mathcal{N}$ includes all the nuisance parameters. 
For SNe Ia, the nuisance parameters are the B-band absolute
magnitude $M_B$, the magnitude offset $\Delta_{M}$ for more massive
host galaxies, and the light-curve parameters $\alpha$ and
$\beta$~\citep{2014A&A...568A..22B};
for \emph{Planck}, the nuisance parameter is $A_{Planck}$,  
the absolute map calibration~\citep{2016A&A...594A..13P,2016A&A...594A...8P}.
The EoS parameters $\{w_i\}$ are assigned according to the scheme
described in section~\ref{subsec:method} with $a_\mathrm{min}=0.286$
($z_\mathrm{max}=2.5$).
Since the parameters $A_s$, $n_s$, $\tau$, and $A_{Planck}$ do
not affect the CMB distance prior, they are removed from the parameter
set when JD16$^*$ is used for reconstruction.

\begin{figure}[t!] 
\centering
\includegraphics[width=0.6\textwidth]{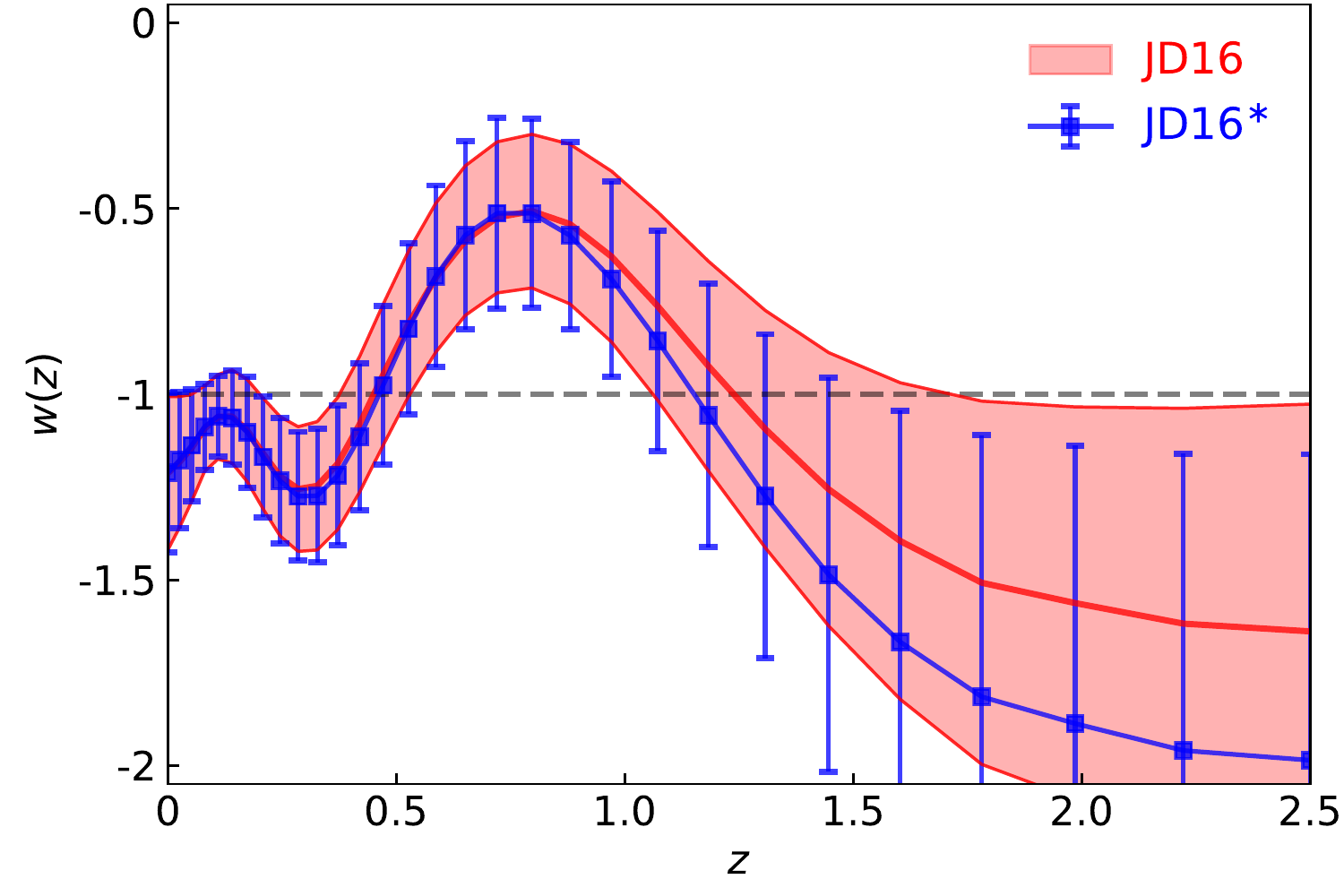}
\caption{Dark energy EoS reconstructed from JD16 (red band)
  and JD16$^*$ (blue squares). The band and error bars
  indicate the marginalized $1\sigma$ error.
}
\label{fig:eos1}
\end{figure}

The results of reconstruction are shown in figure~\ref{fig:eos1}.
The dark energy EoS from JD16$^*$ (blue squares) traces that from
JD16 (red band) very well, and they display marked resemblance to 
those in refs.~\citep{2017NatAs...1..627Z,2018ApJ...857....9D}.
Figure~\ref{fig:eos1} demonstrates that the CMB distance prior, which
greatly reduces the amount of computation, can replace the full CMB
power spectra for non-parametric reconstruction of the dark energy
EoS.

\section{Tests in the concordance cosmological model}
\label{sec:tests}

As with any parameter estimation problem, one must ensure the
fidelity of the reconstructed dark energy EoS.
Specifically, the actual EoS should be the only source of any
statistically significant feature in the reconstructed EoS.
This section performs tests on mock data generated for the flat
$\Lambda\rm{CDM}$ universe to see whether characteristics
of dynamical dark energy can appear in the reconstructed EoS
at high significance. These tests are included in the ClassMC
package as examples.

\subsection{Reconstruction using mock data of existing observations}
\label{sec:ljd16}
\subsubsection{Mock datasets}

The JD16$^*$ dataset is used as a template to simulate an ensemble
of 1000 mock datasets for the $\Lambda\rm{CDM}$ model
(\LJD{} datasets). Details are as follows.

\begin{itemize}
\item
  The fiducial cosmology is specified by the \emph{Planck} 2018
  base-$\Lambda\rm{CDM}$ cosmological
  parameters~\citep[][table 1]{2018arXiv180706209P}. In particular,
  $H_0=67.32~\rm{km~s^{-1} Mpc^{-1}}$, $\Omega_b h^2=0.022383$, and 
  $\Omega_c h^2=0.12011$, where $h$ is the reduced Hubble constant.

\item
  The covariance matrix of \LJD{} is the same as that of JD16$^*$, 
  except that the light-curve model parameters $\alpha$ and $\beta$
  in the JLA covariance matrix are fixed to their best-fit
  values~\citep{2013A&A...552A.124B,2014A&A...568A..22B}.

\item
  The redshift of each data point is exactly the same as that in JD16$^*$.
  The fiducial values of the distance, the Hubble constant,
  and the Hubble parameter are given by the fiducial cosmology.
  Random errors are drawn from the multivariate Gaussian
  distribution as defined by the covariance matrix of \LJD{}
  and then added to these fiducial values to form a realization of \LJD{}.

\item
  The covariance matrix of \LJD{} is factorized into
  $\bm{C}_d = \bm{Q}\bm{D}\bm{Q}^{-1}$ with $\bm{Q}$ being an
  orthogonal matrix and $\bm{D}$ being a diagonal matrix.
  A random vector $\bm{v}$ is constructed by drawing each element
  $v_i$ from a zero-mean Gaussian distribution of variance $D_{ii}$.
  The transformation $\Delta\bm{d}=\bm{Q}\bm{v}$ generates the
  random errors that follow the multivariate Gaussian distribution
  as defined by $\bm{C}_d$.  These errors are then added to the
  fiducial values to form a realization of \LJD{}.

\item
  For the SN Ia sample, the B-band absolute magnitude $M_B$
  assumes the fiducial value of $-19.3$, and the magnitude offset $\Delta_{M}$ is set to zero.
  
\item
  The fiducial values of the quantities constituting the CMB distance
  prior are calculated using the fitting formulae from
  refs.~\citep{2007PhRvD..76j3533W,2009ApJS..180..330K} (see
  appendix~\ref{app:cmb_dist}).

\end{itemize}

Altogether, there are 758 degrees of freedom for estimating the set of
parameters \{$H_0$, $\Omega_b$, $\Omega_c$, $w_0$, \dots, $w_{29}$, $M_B$\}
from each \LJD{} dataset. For convenience, we refer to the aforementioned
multivariate Gaussian distribution of the data as $\mathcal{G}(\bm{C}_d)$.

\subsubsection{Examples of the reconstructed EoS}
\label{subsec:result}

\begin{figure}
\centering
\includegraphics[width=0.495\textwidth]{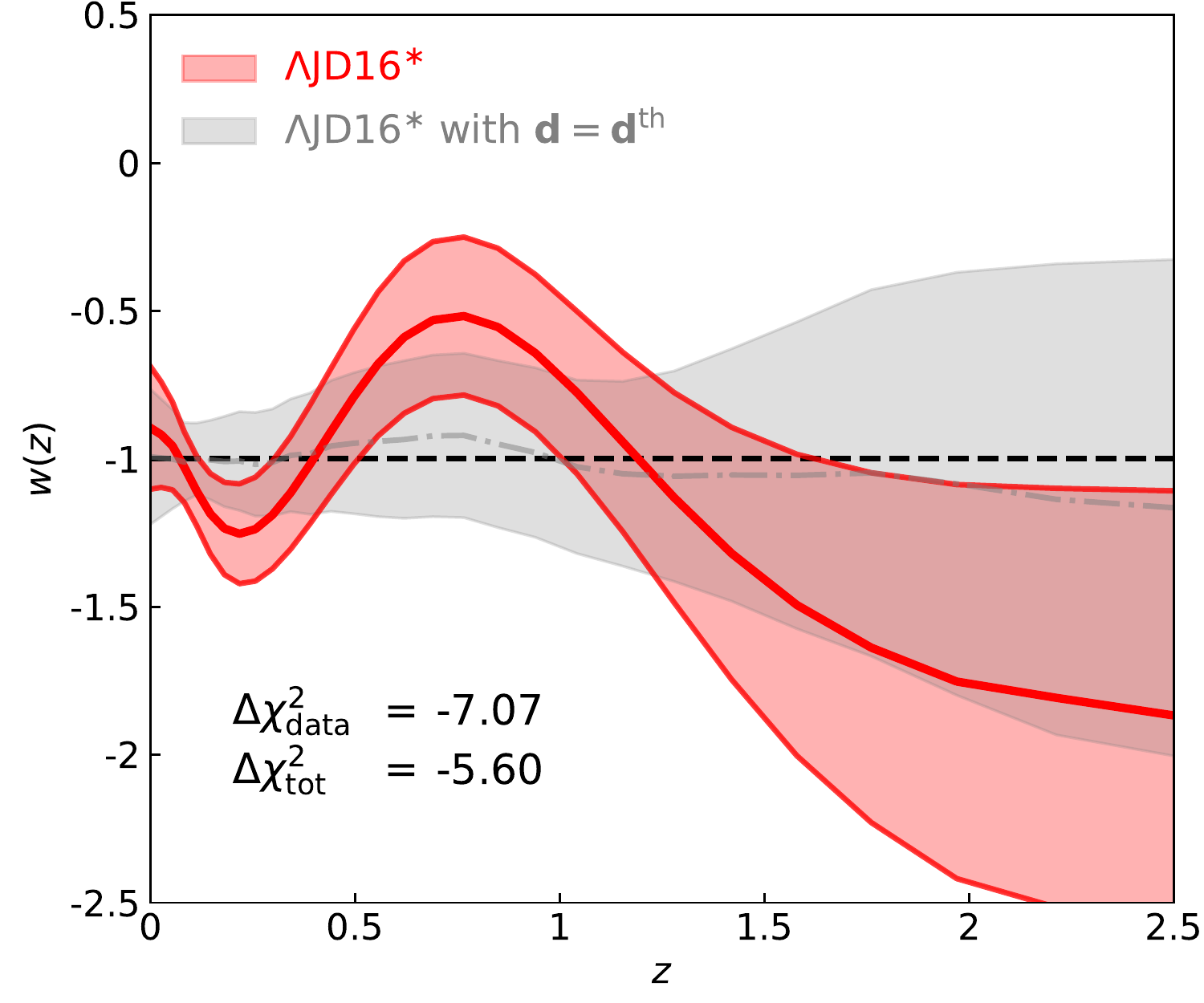}
\includegraphics[width=0.495\textwidth]{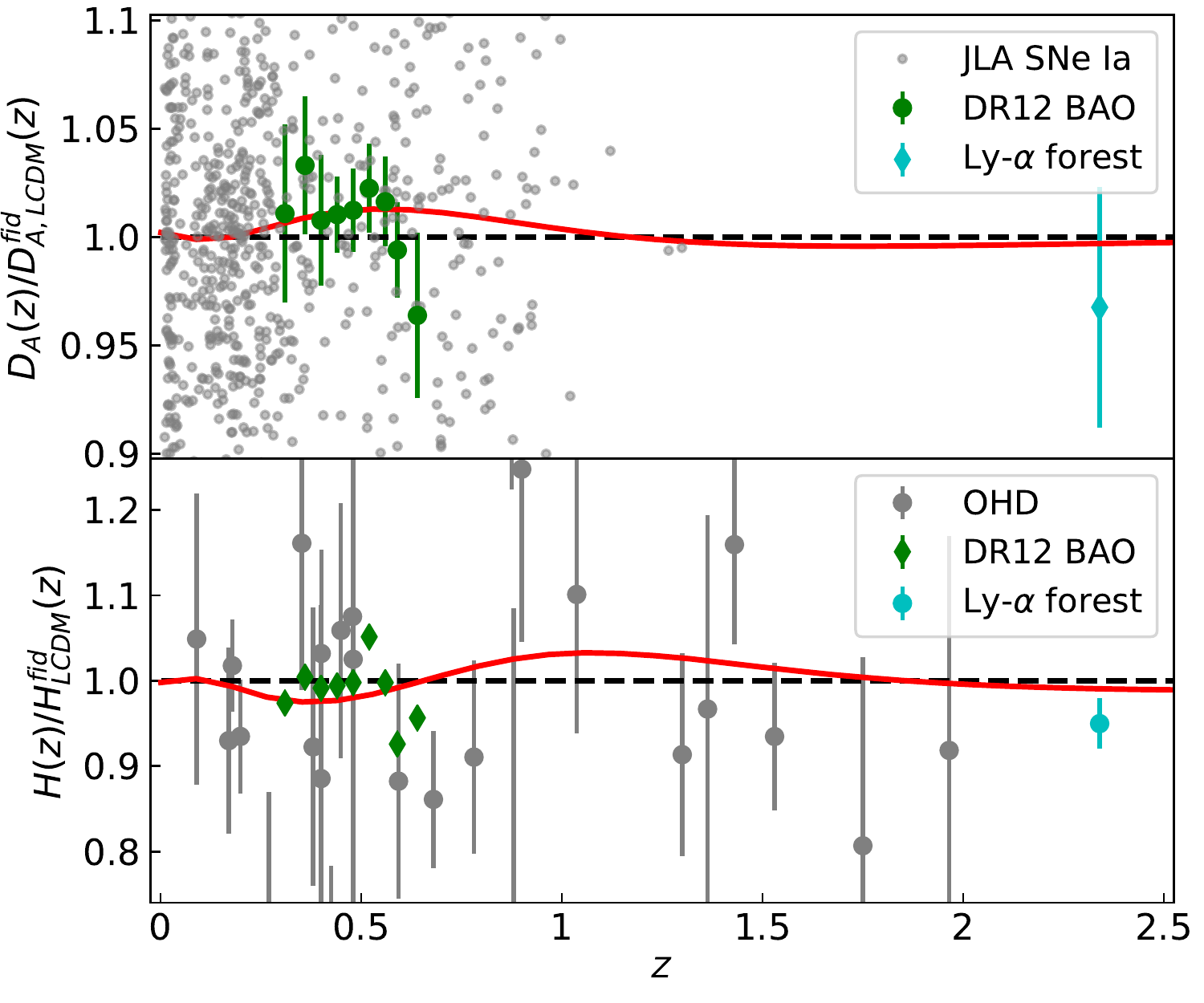}
\caption{
  \emph{Left panel}: Dark energy EoS (red band) reconstructed
  from a realization of \LJD{}.
  The grey band is the result with all the data set
  exactly by the fiducial $\Lambda$CDM model.
  \emph{Right panel}: Mock data of the angular diameter distance
  and the Hubble parameter scaled by corresponding values of the
  fiducial $\Lambda$CDM model. For SNe Ia, the angular diameter
  distances are converted from their luminosity distances and the
  uncertainties are omitted for clarity. The red solid lines are the
  predictions of the best-fit $\tilde{w}$CDM model.
}
\label{fig:eos_mock}
\end{figure}

An example of the reconstructed dark energy EoS is shown in the left
panel of figure~\ref{fig:eos_mock} (red band).
The corresponding scaled mock distance and Hubble parameter data
of JLA SNe Ia, BOSS DR12 BAO, OHD, and Ly$\alpha$ forest BAO are
displayed in the right panel of figure~\ref{fig:eos_mock}, along with
the values calculated using the best-fit $\tilde{w}$CDM model
(red solid lines).
The difference between the best-fit $\chi_{\rm data, BF}^2$ and that of
the $\Lambda$CDM model $\chi_{\rm data, \Lambda}^2$ is
$\Delta\chi_{\rm data}^2=-7.07$ and would
have been interpreted as a $2.7\sigma$ ($|\Delta\chi_{\rm data}^2|^{1/2}$) 
preference for the $\tilde{w}$CDM model over the $\Lambda$CDM
model~\citep{2017NatAs...1..627Z}. 
Although $\chi_{\rm data}^2$ does not include the contribution from the
prior directly, the difference $\Delta\chi_{\rm data}^2$ is affected by
the prior through the selection of the best-fit model.

The correlation prior is an integral part of the parameter
estimation process that produces the reconstructed EoS. It is also
appropriate to take the prior contribution into account and compare
the goodness-of-fit between the two models based on $\chi_{\rm tot}^2$. 
The resulting $\Delta\chi_{\rm tot}^2=-5.60$ is weaker than
$\Delta\chi_{\rm data}^2$ as it should be, and the preference
for the $\tilde{w}$CDM model becomes less significant.
This particular example is selected for its resemblance to the result
in ref.~\citep{2017NatAs...1..627Z}.
It suggests that the usual statistical interpretation of
$\Delta\chi_{\rm data}^2$ (or even $\Delta\chi_{\rm tot}^2$) may not be
sufficient as evidence for dynamical dark energy from
non-parametric reconstruction of the EoS with the correlation prior.

\begin{figure}
\centering
\includegraphics[width=\textwidth]{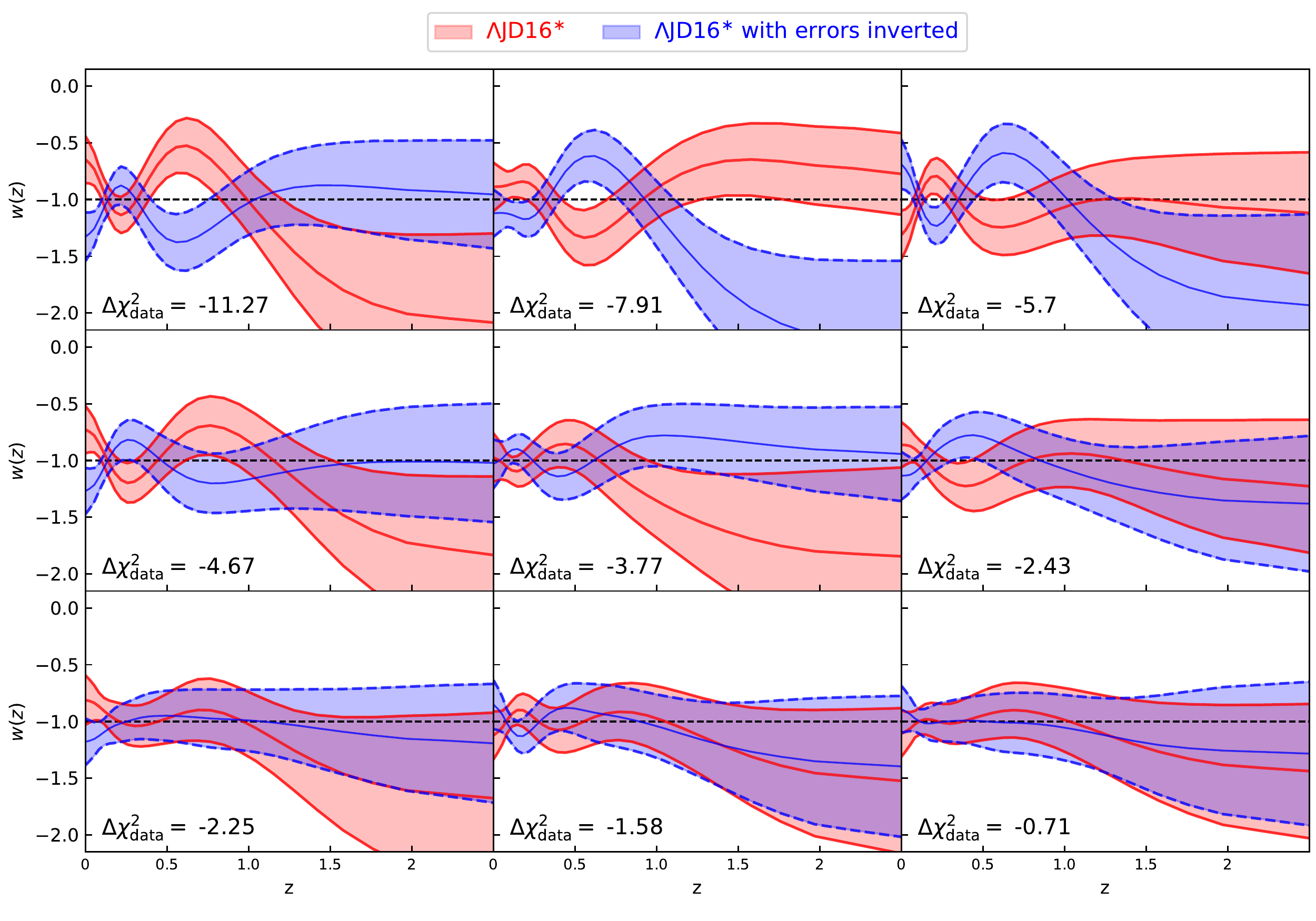}
\caption{
  Randomly selected examples of the dark energy EoS reconstructed
  from \LJD{} datasets. The red bands are the original results,
  and the blue bands are those with all the random errors of the
  data inverted. The panels are ordered by
  $\Delta \chi_{\rm data}^2$ of the original results.
}
\label{fig:eos_list}
\end{figure}

Since the correlation prior is designed to suppress
fast-varying EoS modes not to create any modes, the survived
oscillating feature must be sourced in the mock data.
The underlying $\Lambda$CDM cosmology of \LJD{} does
not produce the signal either.
This leaves the statistical errors of the data the only suspect. 
The grey band in the left panel of figure~\ref{fig:eos_mock}
is the result of a test in which all the mock data are set to
their fiducial values, i.e., $\bm{d}=\bm{d}^\mathrm{th}$.
In this way, the fiducial $\Lambda$CDM model is the best-fit with
$\chi_{\rm tot}^2=0$ by construction, and the result is
highly consistent with $w=-1$ as expected.
It is evident from this test that the errors in the data are the
source of the oscillating feature in the EoS reconstructed from
the \LJD{} dataset.

Figure~\ref{fig:eos_list} displays nine randomly selected examples
of the reconstructed EoS (red bands). In seven of the nine examples,
the recovered EoS deviates from the fiducial model $w=-1$ by
more than $1\sigma$ at least once over the whole redshift range.
Because of the overall constraint from the distance to the last
scattering surface, it is unlikely for the recovered dark energy
EoS to always stay above or below $w=-1$.
Therefore, a departure of the EoS in one bin from
$w=-1$ often causes the EoS in other bin(s) to deviate in the
opposite direction for compensation, producing a wiggle in the EoS.

For each \LJD{} dataset in figure~\ref{fig:eos_list}, we also create a
conjugate dataset by inverting the signs of the errors of the data
in the original dataset. The fiducial $\Lambda$CDM model has
exactly the same $\chi_{\rm tot}^2$ with both datasets.
Interestingly, the original and conjugate results are
roughly symmetric around $w=-1$ at $z \lesssim 1$, 
which demonstrates that
the signature of dynamical dark energy in the EoS reconstructed from
\LJD{} indeed originates from the random errors of the data.

\begin{figure}[t!]
\centering
\includegraphics[width=\textwidth]{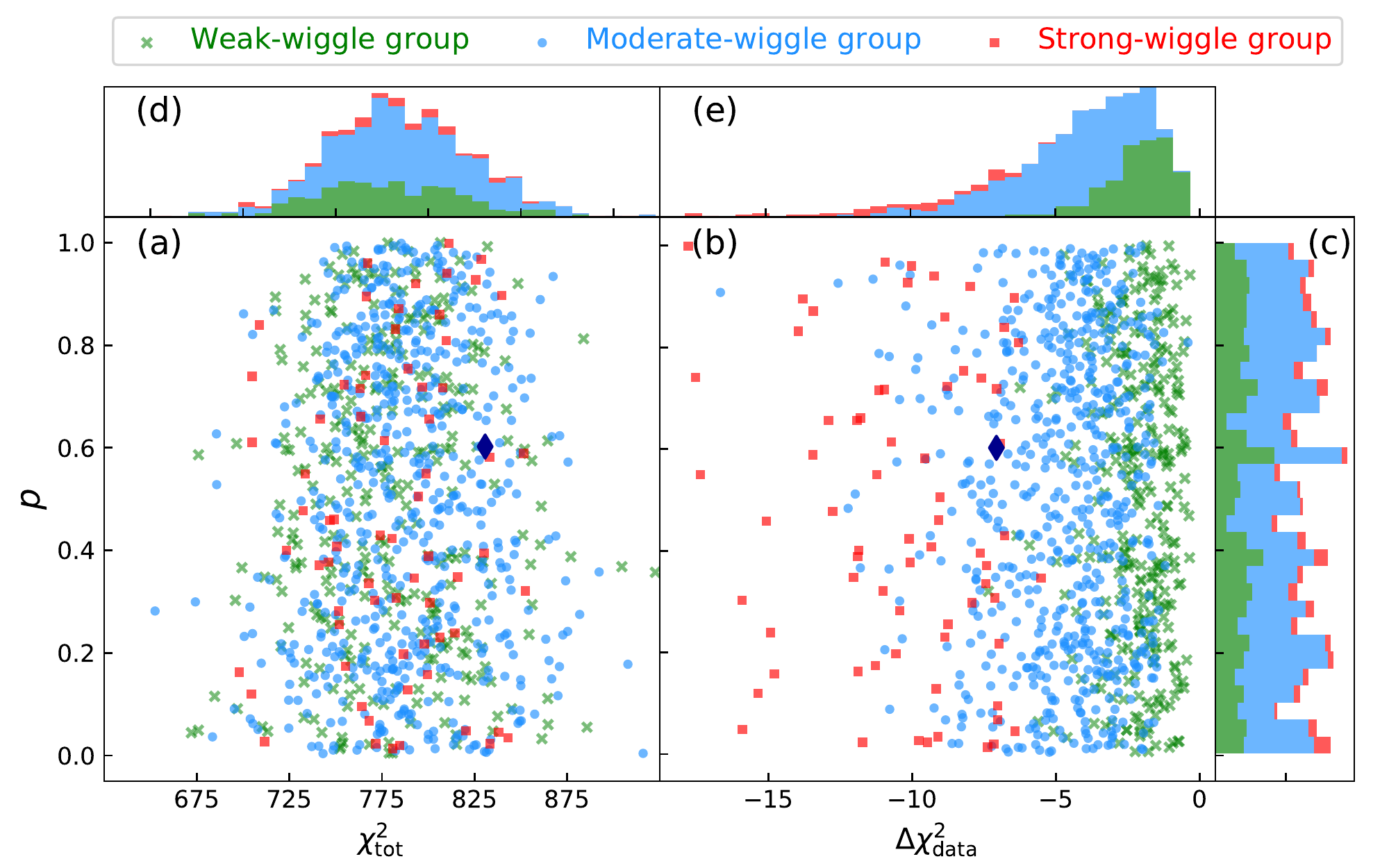}
\caption{
Distributions of the best-fit $\chi_\mathrm{tot}^2$, 
$\Delta \chi_\mathrm{data}^2$, and $p$-values of the 1000 \LJD{} datasets.
\emph{Panels (a) and (b)}:
Green crosses, blue dots, and red squares represent, respectively,
the weak-wiggle group, the moderate-wiggle group, and the
strong-wiggle group. See text for definitions of the three groups.
The blue diamond corresponds to the \LJD{} result in
figure~\ref{fig:eos_mock}, which belongs to the moderate-wiggle group. 
The color-coding is the same for the other panels.
\emph{Panel (c)}: Stacked histogram of the KS-test $p$-values of
the mock datasets.
\emph{Panels (d) and (e)}: Stacked histograms of the best-fit
$\chi_\mathrm{tot}^2$ and $\Delta \chi_\mathrm{data}^2$.
}
\label{fig:kstest}
\end{figure}

\subsubsection{Statistics of the ensemble}

Figure~\ref{fig:kstest} examines the reconstructed dark energy EoS
in terms of the distributions of the best-fit $\chi_\mathrm{tot}^2$,
$\Delta \chi_\mathrm{data}^2$ with respect to the $\Lambda$CDM model,
and $p$-value of the Kolmogorov-Smirnov test
(KS test) for each \LJD{} dataset. 
The $p$-value provides an estimate of the probability of the data
following a particular distribution.
Although the mock datasets are drawn from the multivariate
Gaussian distribution $\mathcal{G}(\bm{C}_d)$, there is a finite
probability for some datasets to appear incompatible with 
$\mathcal{G}(\bm{C}_d)$ at high significance and produce low
$p$-values.
Therefore, we take the $p$-value as a measure of apparent
departure from $\mathcal{G}(\bm{C}_d)$ and check if it is related
to the occurrence of artificial EoS features. 

In practice, we rotate and rescale the error vector 
$\Delta \tilde{\bm{d}}= \bm{C}_d^{-1/2} (\bm{d}-\bm{d}^\mathrm{th})$,
so that the resulting elements $\Delta \tilde{d}_i$ all follow
the unit-variance Gaussian distribution.
The KS test is then performed on the set of $\Delta \tilde{d}_i$
of each mock dataset against the unit-variance Gaussian distribution.
The results are displayed intuitively in three groups according to
the behavior of the best-fit EoS:
(1) the weak-wiggle group, which has an EoS always less than
$1\sigma$ away from $w=-1$, i.e., $|1+w_i|<\sigma_{w_i}$ in all bins;
(2) the moderate-wiggle group, which has $|1+w_i|\ge\sigma_{w_i}$
in at least one bin and $|1+w_i|<2\sigma_{w_i}$ in all bins; and
(3) the strong-wiggle group, which has $|1+w_i| \ge 2\sigma_{w_i}$
in at least one bin.

It is seen from panels (a) and (b) of figure~\ref{fig:kstest} that
the $p$-value is uncorrelated with the best-fit $\chi_\mathrm{tot}^2$
and $\Delta\chi_\mathrm{data}^2$.
Highly significant spurious signals of dynamical dark energy
(large $|\Delta\chi_\mathrm{data}^2|$) can be obtained from
\LJD{} datasets over the full range of $p$-values from appearing
incompatible with $\mathcal{G}(\bm{C}_d)$ ($p$-value $\sim 0$) to
a perfect match ($p$-value $\sim 1$).
The example EoS in figure~\ref{fig:eos_mock} as marked by a
blue diamond in figure~\ref{fig:kstest} is not an extreme case in terms
of $p$-value, $\chi_\mathrm{tot}^2$, and $\Delta\chi_\mathrm{data}^2$.
The histogram of the $p$-value in panel (c) is roughly flat as expected. 
Moreover, none of the $p$-value histogram components of the three
EoS groups displays any clear trend with the $p$-value itself.

Histograms of $\chi_\mathrm{tot}^2$ and $\Delta\chi_\mathrm{data}^2$
are shown, respectively, in panels (d) and (e) of figure~\ref{fig:kstest}.
The $\chi_\mathrm{tot}^2$ distribution peaks around 775 with
a maximum $\chi_\mathrm{tot}^2$ of 923.6, which is normal for
parameter estimation with 758 degrees of freedom.
The three EoS groups behave similarly in the $\chi_\mathrm{tot}^2$
histogram but do differ in the $\Delta \chi_\mathrm{data}^2$ histogram
in a way that is consistent with their intuitive classification:
the weak-wiggle group (312 cases out of 1000 realizations) generally
does not give a strong preference for dynamical dark energy, i.e.,
mostly $|\Delta \chi_\mathrm{data}^2|<5$, and dominates the whole
distribution at $|\Delta \chi_\mathrm{data}^2|\lesssim 2.5$;
the strong-wiggle group (69 cases)
has $|\Delta \chi_\mathrm{data}^2|>5$ and is the main
contributor at $|\Delta \chi_\mathrm{data}^2|\gtrsim 9$;
and the moderate-wiggle group (619 cases) is an intermediate between
the other two groups. Of the 1000 mock datasets, there are 72 cases
preferring the $\tilde{w}$CDM model to the $\Lambda$CDM model at
$\ge 3\sigma$ level ($|\Delta\chi_\mathrm{data}^2|\ge9$).

\subsection{Reconstruction using more data from future surveys}

It is often assumed that the effect of the prior weakens as more data
becomes available. Indeed, upcoming surveys, 
such as LSST, DESI, \emph{Euclid} and \emph{WFIRST},  will provide
unprecedentedly powerful data for dark energy studies.
Hence, we test here whether future data could help
eliminate artificial EoS features.

Using the \LJD{} data model of the existing observations as a baseline,
we add the expected \emph{WFIRST} SN Ia sample and replace the original
mock BAO data with those anticipated from DESI, \emph{Euclid}, and
\emph{WFIRST} to form the future \LJDF{} data model. 
For each \LJDF{} dataset, mock data of future BAO
and SN Ia observations are generated randomly according to the
following description.

\begin{figure}[t!]
\centering
\includegraphics[width=0.66\textwidth]{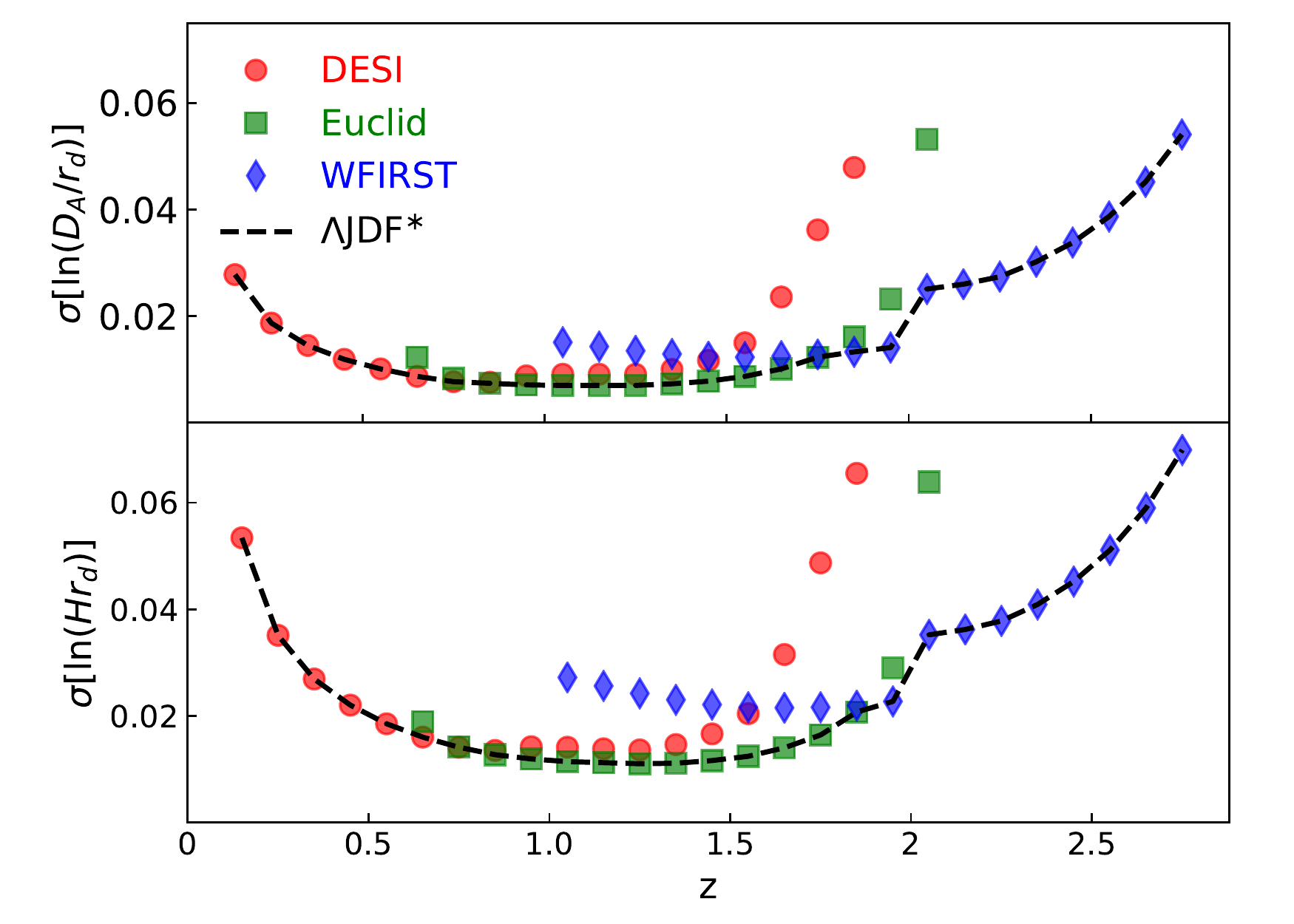}
\caption{
  \emph{Upper panel}: Expected fractional errors of the angular
  diameter distance $D_A(z)$ from future BAO surveys~\citep{Font_Ribera_2014}.
  The lower envelope of the distance errors is used for \LJDF{} datasets.
  \emph{Lower panel}: Same as the upper panel but for the Hubble
  parameter $H(z)$. 
}
\label{fig:forecast_err}
\end{figure}

\begin{itemize}
\item
  The expected errors of the angular diameter distance $D_A(z)$ and
  the Hubble parameter $H(z)$ from DESI, \emph{Euclid}, and
  \emph{WFIRST} are given by ref.~\citep{Font_Ribera_2014} and are
  shown in figure~\ref{fig:forecast_err}.
  For simplicity, we do not attempt to combine all the BAO data in
  a statistically consistent way, which would involve properly
  treating overlapping volumes and overlapping samples between the
  surveys as well as correlations between the BAO data.
  Instead, we keep only the data point of the smallest error at each
  redshift and neglect correlations between the BAO data.

\item
  The redshift distribution and error model of the \emph{WFIRST}
  SNe Ia data are adopted from ref.~\citep{2015arXiv150303757S}.
  There is a total of 2725 SNe Ia in the redshift range $0.1<z<1.7$.
  The errors of the distance modulus are assumed to be independent
  between different SNe Ia and follow a redshift-dependent Gaussian
  distribution with a total variance
\begin{eqnarray}
  \sigma^2_{\mu} = \sigma^2_{\rm int} + \sigma^2_{\rm meas} +
  \sigma^2_{\rm lens} + \sigma^2_{\rm sys},
\end{eqnarray}
where $\sigma_{\rm int}=0.09$ is the intrinsic spread of the SN Ia peak
absolute magnitude (after correction for the light-curve shape and
spectral properties), $\sigma_{\rm meas}=0.08$ is the photometry error,
$\sigma_{\rm lens}=0.07 z$ accounts for lensing magnification, and
$\sigma_{\rm sys}=0.02/(1+z)/1.8$ is attributed to potential systematics.
\end{itemize}

\begin{figure}[t!]
\centering
\includegraphics[width=\textwidth]{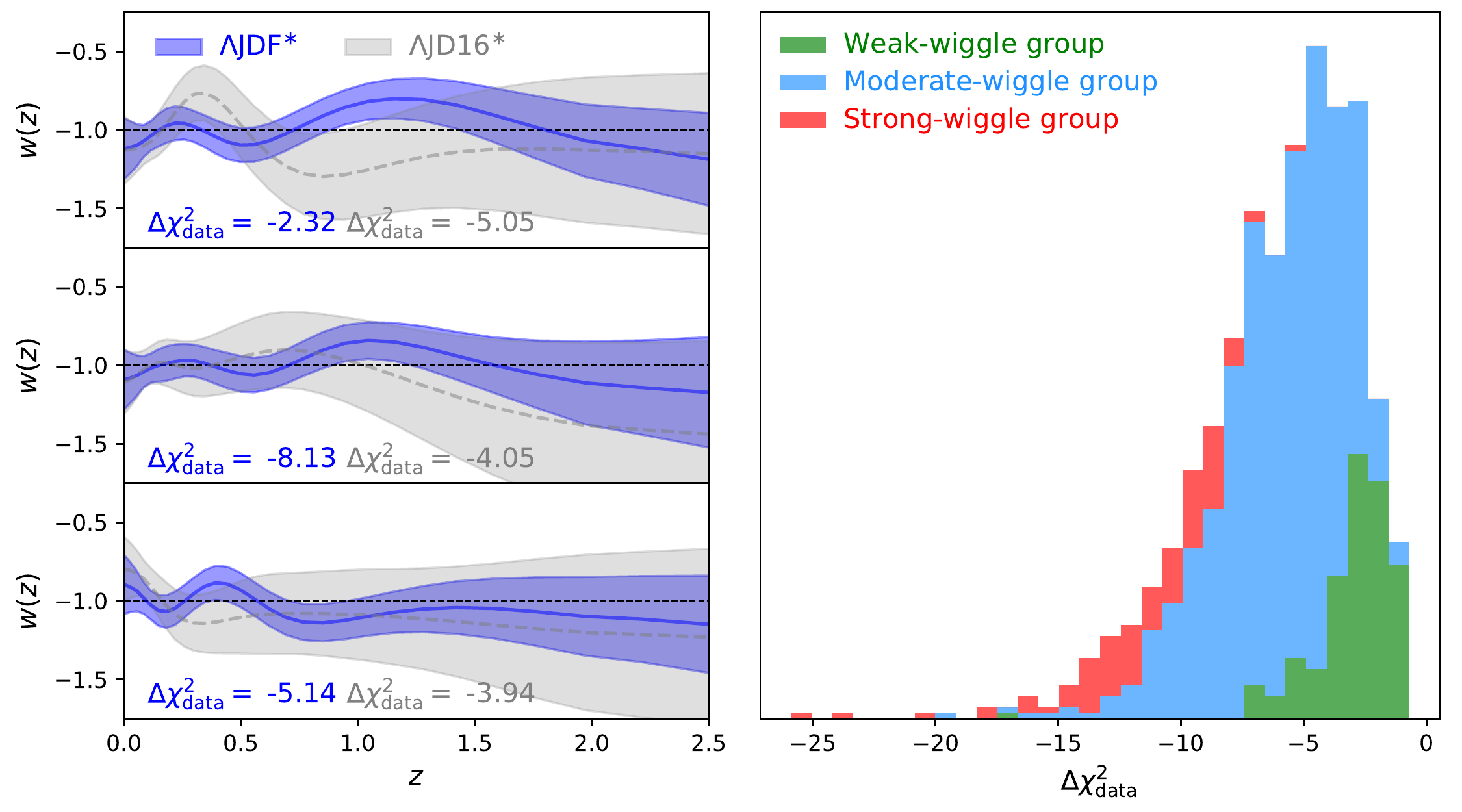}
\caption{
  Results of dark energy EoS reconstruction from the \LJDF{} datasets.
  \emph{Left panel}: Three randomly selected examples of the
  reconstructed EoS (blue) along with their
  corresponding \LJD{} results (grey).
  \emph{Right panel}: Stacked histogram of the best-fit
  $\Delta\chi_\mathrm{data}^2$.
  The color-coding is the same as that in figure~\ref{fig:kstest}.
}
\label{fig:forecast_w}
\end{figure}

The parameters to be estimated from each \LJDF{} dataset are the
same as those in section~\ref{sec:ljd16}, and there are 3515
degrees of freedom in total. One may worry that ignoring the
correlations between the data can lead to unduly tight constraints
on the EoS.
Since we only need to check the significance of spurious features
in the EoS that is reconstructed with more data, ignoring the
correlations does not affect the result qualitatively.

The left panel of figure~\ref{fig:forecast_w} displays three randomly
selected examples of the reconstructed dark energy EoS from \LJDF{}
datasets (blue bands) along with the ones from corresponding \LJD{}
datasets (grey bands).
The reduction of the uncertainties of the reconstructed EoS with future
SNe Ia and BAO data is quite notable. However, the preference for
dynamical dark energy, as quantified by $\Delta \chi_\mathrm{data}^2$,
can become even stronger for some \LJDF{} datasets. In fact,
the $\Delta \chi_\mathrm{data}^2$ histogram of the \LJDF{} results
in figure~\ref{fig:forecast_w} is shifted towards more negative values
with respect to that of the \LJD{} results in figure~\ref{fig:kstest}.
On average, the \LJDF{} datasets
increase $|\Delta \chi_\mathrm{data}^2|$ by 1.72 over that of the \LJD{} datasets.
The occurrence of the moderate-wiggle group and the strong-wiggle group
is also increased, respectively, from 619 cases and 69 cases
to 724 cases and 100 cases out of 1000 realizations.
There are now 171 cases reaching $|\Delta \chi_\mathrm{data}^2|\ge 9$,
more than doubling those of the \LJD{} results.
Hence, there is no guarantee that non-parametric reconstruction of
the dark energy EoS with the correlation prior can eliminate spurious
EoS features with more data.

\section{Summary}
\label{sec:summary}

We have carried out a number of tests on non-parametric
reconstruction of the dark energy EoS with the correlation prior
as proposed by refs.~\citep{2012JCAP...02..048C}.
While we obtain an oscillating EoS similar to that in
ref.~\citep{2017NatAs...1..627Z} using largely the same set of
observational data, highly resemblant results can also be
obtained from \LJD{} mock datasets of existing observations that
are generated under the flat $\Lambda$CDM model.
The \LJDF{} mocks that include future SN Ia and BAO data
reduce the EoS uncertainties considerably, though they
actually increase the occurrence of significant spurious features
in the reconstructed EoS. 
For instance, the fraction of realizations that prefers
dynamical dark energy to the cosmological constant at $\ge3\sigma$
level is $7.2\%$ with the \LJD{} mocks and is boosted to
$17.1\%$ with the \LJDF{} mocks.
It is evident that, with the correlation prior, adding more data
does not necessarily recover the true EoS.

Figure~\ref{fig:eos_list} demonstrates that the source of the
oscillating EoS features can be traced to random errors in the data. 
However, random errors always exist and can produce all sorts of
modes in the EoS. Once the poorly determined high-frequency EoS
modes are suppressed by the correlation prior, the EoS uncertainties
become small enough to give a good chance for the surviving
slowly-oscillating modes to stand out at high significance.
In other words, even though those poorly determined high-frequency
EoS modes are undesirable, they are part of the complete
statistical description of the reconstructed EoS without which one
would severely underestimate the uncertainties of the EoS.

Since random errors in the data can generate time-varying EoS
modes that behave like dynamical dark energy, it would be difficult
to draw a conclusion based on existing results.
As such, more studies of the correlation prior and reconstruction
method are needed to accurately determine the dark energy EoS and
test it against theoretical models.

\acknowledgments

HZ acknowledges support from the National Science Foundation of
China grant No.~11721303, the National Key R\&D Program of China
grant No.~2016YFB1000605, and China Manned Space Program through
the Space Application System.
This research of YKEC  has been supported in parts 
by the NSF China under Contract No.~11775110, and No.~11690034.
YKEC also acknowledges the European Union's Horizon 2020 research
and innovation programme (RISE) under the Marie Sk\'lodowska-Curie
grant agreement No.~644121, and  the Priority Academic Program
Development for Jiangsu Higher Education Institutions (PAPD).

\appendix

\section{Fitting formulae for the CMB distance prior}
\label{app:cmb_dist}

The CMB distance prior consists of the acoustic scale $l_{\rm A}$,
the shift parameter $R$, and the photon decoupling redshift $z_*$.
These quantities are calculated using the fitting formulae from
refs.~\citep{2007PhRvD..76j3533W,2009ApJS..180..330K}:
\begin{eqnarray}
l_{\rm A} &=& (1+z_*)\frac{\pi D_{\rm A}(z_*)}{r_s(z_*)},\\
R &=& \frac{\sqrt{(\Omega_{\rm b}+\Omega_{\rm c})H_0^2}}{c}(1+z_*)D_{\rm A}(z_*),\\
z_* &=& 1048\left[1+0.00124(\Omega_{\rm b}h^2)\right]^{-0.738}
\left[1+g_1(\Omega_{\rm b}h^2+\Omega_{\rm c}h^2)^{g_2}\right],
\end{eqnarray}
where 
\begin{eqnarray}
g_1 &=& \frac{0.0783(\Omega_{\rm b}h^2)^{-0.238}}{1+39.5(\Omega_{\rm b}h^2)^{0.763}},\\
g_2 &=& \frac{0.560}{1+21.1(\Omega_{\rm b}h^2)^{1.81}}.
\end{eqnarray}
The angular diameter distance $D_{\rm A}(z)$ and the sound horizon $r_{\rm s}(z)$ are defined as:
\begin{eqnarray}
D_{\rm A}(z) &=& \frac{c}{(1+z)} \int_{0}^{z} \frac{dz'}{H(z')}, \\
r_{\rm s}(z) &=& \frac{c}{\sqrt{3}} \int_{0}^{(1+z)^{-1}}\frac{da}{a^2H(a)\sqrt{1+(3\Omega_{\rm b}/4\Omega_{\rm \gamma})a}},
\end{eqnarray}
where $\Omega_\gamma$ is the radiation density parameter.

\bibliographystyle{JHEP}
\bibliography{EoS}

\end{document}